\documentclass[aps,twocolumn,superscriptaddress,preprintnumbers]{revtex4}

\usepackage[latin9]{inputenc}
\usepackage{color}
\usepackage{verbatim}
\usepackage{amsmath}
\usepackage{amssymb}
\usepackage{graphicx}
\usepackage{esint}
\usepackage{hyperref}

\makeatletter
\@ifundefined{textcolor}{}
{%
 \definecolor{BLACK}{gray}{0}
 \definecolor{WHITE}{gray}{1}
 \definecolor{RED}{rgb}{1,0,0}
 \definecolor{GREEN}{rgb}{0,1,0}
 \definecolor{BLUE}{rgb}{0,0,1}
 \definecolor{CYAN}{cmyk}{1,0,0,0}
 \definecolor{MAGENTA}{cmyk}{0,1,0,0}
 \definecolor{YELLOW}{cmyk}{0,0,1,0}
}

\usepackage{mathrsfs}

\draft
\makeatother

\begin{document}

\title{Topological Spin Texture in a Quantum Anomalous Hall Insulator}

\author{Jiansheng Wu}
\affiliation{Department of Physics, Hong Kong University of Science and Technology, Clear Water Bay, Hong Kong, China}
\author{Jie Liu}
\affiliation{Department of Physics, Hong Kong University of Science and Technology, Clear Water Bay, Hong Kong, China}
\author{Xiong-Jun Liu \footnote{Corresponding author: xiongjunliu@pku.edu.cn}}
\affiliation{Department of Physics, Hong Kong University of Science and Technology, Clear Water Bay, Hong Kong, China}
\affiliation{International Center for Quantum Materials and School of Physics, Peking University, Beijing 100871, China}
\affiliation{Department of Physics, Massachusetts Institute of Technology, Cambridge, Massachusetts 02139, USA}

\begin{abstract}
The quantum anomalous Hall (QAH) effect has been recently discovered in experiment using thin-film topological insulator with ferromagnetic ordering and strong spin-orbit coupling. Here we investigate the spin degree of freedom of a QAH insulator and uncover a fundamental phenomenon that the edge states exhibit topologically stable spin texture in the boundary when a chiral-like symmetry is present. This result shows that edge states are chiral in both the orbital and spin degrees of freedom, and the chiral edge spin texture corresponds to the bulk topological states of the QAH insulator. We also study the potential applications of the edge spin texture in designing topological-state-based spin devices which might be applicable to future spintronic technologies.
\end{abstract}
\date{\today }
\maketitle



Discovery of the quantum Hall effect in 1980s brought about a fundamental concept, topological state of quantum matter, to condensed matter physics~\cite{QHE,FQHE}. In the quantum Hall effect an external magnetic field drives the electrons to fill in discrete Landau levels. This leads to a gap in the two-dimensional (2D) bulk, while the boundary exhibits 1D chiral gapless edge states. The Hall conductance in the integer quantum Hall effect is quantized by Chern numbers, which are integers and characterize the topology of the system~\cite{TKNN}. The topological interpretation of the quantized Hall conductance implies that to obtain a quantum Hall state does not necessitate an external magnetic field. The theoretical idea for quantum Hall effect without Landau levels, i.e. the quantum anomalous Hall (QAH) effect, was first introduced by Haldane in a honeycomb lattice over two decades ago~\cite{Haldane}. The recent interests in this topological state have been revived due to the great developments in the time-reversal-invariant topological insulators~\cite{TI1,TI2}, with many new theoretical proposals having been introduced~\cite{QAHE1,QAHE2,Congjun2008,FangZhong1,QAHE3,QAHE4,Chaoxing}. Importantly, following the proposal in Ref.~\cite{FangZhong1}, the QAH state has been detected in a recent experiment using ferromagnetic (FM) thin-film topological insulator by observing the quantized Hall conductance~\cite{Qikun}.

While in QAH effect the spin is generically not conserved, the recent experimental realization based on strongly spin-orbit (SO) coupled materials~\cite{Qikun} motivates us to explore the spin degree of freedom of such QAH insulators. We show surprisingly that the edge states exhibit topologically stable spin texture in the boundary when a chiral-like symmetry is present, and such chiral edge spin texture corresponds to the bulk topological states of the QAH insulator. The potential applications of the edge spin texture are studied.

The minimal model for the QAH insulator is described by a two-band Hamiltonian $H=\sum_{\bold k}\psi_{\bold k}^\dag{\cal H}(\bold k)\psi_{\bold k}$, where the spin basis $\psi_{\bold k}=(c_{\bold k\uparrow},c_{\bold k\downarrow})^T$. Around the $\Gamma$ point ${\cal H}(\bold k)$ takes the simple $(2+1)$D Dirac form
\begin{eqnarray}\label{eqn:Hamiltonian1}
{\cal H}(\bold k)={\left[
\begin{matrix}
m_z+2B(k_x^2+k_y^2)  &  2A_1k_y+i2A_2k_x\\
        2A_1k_y-i2A_2k_x & -m_z-2B(k_x^2+k_y^2)\end{matrix} \right]},
\end{eqnarray}
where for the realization with thin-film FM topological insulators~\cite{FangZhong1,Qikun} $m_z$ depends on the Zeeman term induced by the FM order, the $B$-term characterizes the hybridization between top and bottom thin-film surfaces, and $2A_{1,2}$ equal Fermi velocities of the surface Dirac cones of the parent topological insulator~\cite{3DTI1,3DTI2}. The
QAH phase is obtained when $m_zB<0$, with the Chern number $C_1=\mbox{sgn}(m_zA_1A_2)$~\cite{FangZhong1,Qi2008}. In the solid-state experiment, the Cr-doped (Bi$_{1-x}$Sb$_x$)$_2$Te$_3$ was used to achieve the above Hamiltonian and the QAH phase~\cite{Qikun}. On the other hand, this model can also be realized in a square optical lattice~\cite{Liu2013} with SO coupling generated based on the cold atom experiments~\cite{Liu2009,Lin,Wang,MIT}.

\begin{figure*}[t]
\includegraphics[width=1.8\columnwidth]{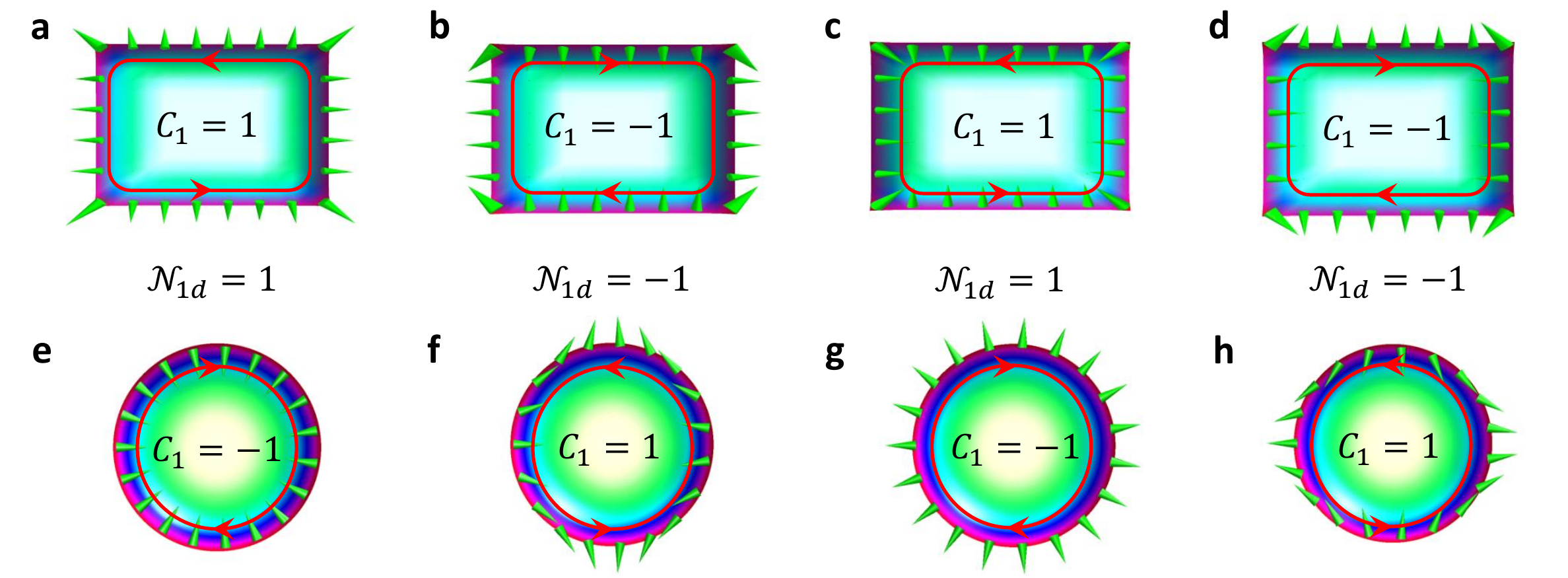}\caption{(a)-(d) The edge spin texture for square geometry of the boundary, with the Zeeman term $m_z>0$ and $B<0$. \small {(e)-(h),} The edge spin texture for circular geometry of the boundary, with $m_z<0$ and $B>0$. For the other parameters, we take that $A_1,A_2>0$ (a,e); $A_1>0, A_2<0$ (b,f); $A_1,A_2<0$ (c,g); and $A_1<0, A_2>0$ (d,h). The topological spin textures give rise to quantized Berry phases if evolving the edge spin one circle along the boundary, which define 1D topological states in the position space of the boundary and characterized by the 1D winding number ${\cal N}_{1d}$. The bulk Chern number $C_1$ corresponds to ${\cal N}_{1d}$ with an additional sign factor $\mbox{sgn}(m_z)$.}
\label{spintexture}
\end{figure*}
In the typical isotropic case $|A_1|=|A_2|$ a chiral-like symmetry $S=\sigma_{\vec n_1}{\cal M}_{\vec n_2}$ emerges in the Hamiltonian, with $\vec n_1\perp\vec n_2$ being arbitrary 2D orthogonal unit vectors in the $x$-$y$ plane. Here $\sigma_{\vec n_1}$ and ${\cal M}_{\vec n_2}$ represent the $\vec n_1$-component of the Pauli matrix acting on the spin space and the spatial reflection along the $\vec n_2$ direction, respectively. The symmetry transforms the fermion operators via $S(c_{\bold ks},c^\dag_{\bold ks})^T=(\sigma_{\vec n_2})_{ss'}(c^\dag_{\bold k's'},c_{\bold k's'})^T$, with $s=\uparrow,\downarrow$, and $\bold k'=(k_{n_1},-k_{n_2})$. One can verify for the second quantization Hamiltonian that $SHS^{\dag}=H$, while the first quantization Hamiltonian satisfies $S{\cal H}(\bold k) S^\dag=-{\cal H}(\bold k)$. Consider two edges normal to $\vec n_1$ axis and we can reexpress the Hamiltonian as a summation of 1D Hamiltonian with fixed momentum $k_{n_2}$ along the edge: $H=\sum_{k_{n_2}}H_{1D}(k_{n_2})$, where
\begin{eqnarray}\label{eqn:Hamiltonian1D1}
H_{1D}(k_{n_2})&=&-\int{dx_{n_1}}\psi_{k_{n_2},x_{n_1}}^{\dag}\biggr\{\bigr[2B\partial^2_{x_{n_1}}+m(k_{n_2})\bigr]\sigma_z\nonumber\\ &&+iA_1\partial_{x_{n_1}}\sigma_{n_2}-A_2k_{n_2}\sigma_{n_1}\biggr\} \psi_{k_{n_2},x_{n_1}},
\end{eqnarray}
with $m(k_{n_2})=m_z+2Bk_{n_2}^2$. The above 1D Hamiltonian satisfies $SH_{1D}(k_{n_2})S^\dag=H_{1D}(-k_{n_2})$, from which we know that $H_{1D}(0)$ describes a 1D topological insulator with chiral symmetry, and its two end states (i.e. edge states with $k_{\vec n_2}=0$ of the 2D system) must be eigenstates of $\sigma_{\vec n_1}$, with the spin oppositely polarized in the opposite edges~\cite{Liu2013AIII}. Furthermore, using the $\bold k\cdot\bold p$ theory one can expand the 1D edge Hamiltonian up to the leading order of momentum and confirm that the edge states with nonzero $k_{\vec n_2}$ are also eigenstates of $\sigma_{\vec n_1}$. This result is can be generalized to the case with a generic closed boundary condition~\cite{SI}. In particular, for the material with a circular boundary, we can rewrite the Hamiltonian in the polar coordinate $(r,\varphi)$ system
\begin{eqnarray}\label{eqn:SIHamiltonian3}
{\cal H}(r,\varphi)&=&\bigr[2B(\frac{1}{r}\partial_rr\partial_r+\frac{1}{r^2}\partial^2_{\varphi})+m_z\bigr]\sigma_z+i\frac{2A_{1}}{r}\sigma_{r}\partial_\varphi\nonumber\\
&& -i2A_{2}\sigma_{\varphi}\partial_{r},
\end{eqnarray}
where $\sigma_r=\cos\varphi\sigma_x+\sin\varphi\sigma_y$ and $\sigma_\varphi=\cos\varphi\sigma_y-\sin\varphi\sigma_x$. The edge state of eigenenergy ${\cal E}_m$ can be described by $|\Phi^{\rm edge}_{m}(r,\varphi)\rangle=|\phi^{\rm edge}_{m}(r,\varphi)\rangle e^{im\varphi}$, with $m$ being integer. It can be verified from the eigen-equation that $S|\phi^{\rm edge}_{m}(r,\varphi)\rangle=|\phi^{\rm edge}_{-m}(r,-\varphi)\rangle$, where $S=\sigma_r{\cal M}_\varphi$ (see Supplementary Material for details~\cite{SI}). On the other hand, the mirror transformation sends ${\cal M}_\varphi|\phi^{\rm edge}_{m}(r,\varphi)\rangle=|\phi^{\rm edge}_{-m}(r,-\varphi)\rangle$. Together with the two relations we reach $\sigma_r|\phi^{\rm edge}_{m}(r,\varphi)\rangle=|\phi^{\rm edge}_{m}(r,\varphi)\rangle$, which shows that all the edge modes are eigenstates of $\sigma_r$ and exhibits spin texture in the boundary.

The numerical results are shown in Fig.~\ref{spintexture} with different signs of $A_{1,2}$, $B$ and $m_z$. We see that the spin of edge states is in-plane polarized, and varies one cycle following the 1D closed path of the boundary. This spin texture shows that the edge states are chiral in both the orbital and spin degrees of freedom. Interestingly, the spin chirality gives a quantized Berry phase for each edge mode after evolving one cycle along the boundary: $\gamma_C=\pm\pi$, and this defines a 1D winding number ${\cal N}_{1d}=\pm1$ which can be verified to correspond to the bulk Chern invariant via $C_1={\cal N}_{1d}\mbox{sgn}(m_z)$. With fixed $\mbox{sgn}(m_z)$, changing the edge spin chirality reverses Chern number of the QAH insulator, while varying both the spin chirality and $\mbox{sgn}(m_z)$ gives the phases with the same $C_1$ [see e.g. Fig.~\ref{spintexture} (a) and (f)]. In Fig.~\ref{spintexture} the chiral spin texture and edge currents are shown in different parameter regimes, and with square (a-d) and circular (e-h) geometries, respectively.

The above study manifests an interesting correspondence between the nontrivial topologies exhibited in the bulk and the boundary. The bulk Chern number is a topological invariant of the first Brillouin zone, which is a 2D closed manifold in the momentum space due to the band gap of the insulator~\cite{TKNN}. However, the 1D boundary is a closed manifold in the position space, but not in the momentum space since the edge modes are gapless. Therefore the 1D edge invariant ${\cal N}_{1d}$ is obtained in the real space rather than in the $\bold k$-space, and the topological edge spin textures can be recognized as 1D real-space topological states. Both the bulk and edge topological states are classified by integers $Z$. Note that a 1D topological state necessitates the symmetry protection~\cite{Schnyder}. The correspondence between the bulk and edge topological phases relies on the chiral-like symmetry as introduced above, albeit the QAH insulator is an intrinsic topological state not depending on symmetry.


While the edge topological state is obtained under the symmetry protection, the chirality enables the topological spin texture to be insensitive to local perturbations which explicitly break the $S$ symmetry. The local perturbation which breaks this symmetry includes the in-plane Zeeman fields, the nonmagnetic and magnetic disorders. In the Supplementary Material we show that the edge spin texture is not affected by the in-plane Zeeman fields without driving phase transition in the bulk, and also insensitive to the local disorder perturbations, even when the disorder strength is comparable with the bulk gap. Actually, like that the orbital chirality of the edge modes prohibits the back scattering, the spin chirality ensures that no scattering occurs between two edge modes with opposite local spin polarizations. On the other hand, in the high-energy regime, due to discrete lattice anisotropy the low-energy Hamiltonian may not apply. For the square lattice model the system only has the $S$-symmetry along $x$ and $y$ directions. This ensures that the edge spin aligns along $x$ ($y$) axis in the edges normal to $\hat e_y$ ($\hat e_x$) direction and far away from sample corners, while around the corners of the square sample the spin polarization of the high-energy edge modes may have a sizable tilt to the perpendicular direction~\cite{SI}.

The edge channel of the QAH insulator can be described by 1D chiral Luttinger liquid~\cite{Halperin,Wen1990}. Furthermore, the above study shows that the edge modes are chiral in both orbital and spin degrees of freedom. As the topological spin texture leads to quantized Berry phases, which can be integrated by Berry's connection, the edge states are governed by the following effective Hamiltonian
\begin{eqnarray}\label{eqn:Luttinger}
H_{\rm edge}=i v_{\rm edge}\int d\tilde x\psi_s^*(\tilde x)\bigr[\partial_{\tilde x}-i{\cal A}_s(\tilde x)\bigr]\psi_s(\tilde x).
\end{eqnarray}
Here $\psi_s$ denotes the orbital part of the edge states, $\tilde x$ is the position parameter along the edge, and the Berry's connection ${\cal A}_s=i\hbar\langle\chi_s(\tilde x|\partial_x|\chi_s(\tilde x)\rangle$, with $|\chi_s(\tilde x)\rangle$ representing the polarized spin degree of freedom. The integral of ${\cal A}_s$ along the 1D boundary gives $\oint d\tilde x{\cal A}_s(\tilde x)={\cal N}_{1d}\pi$. The $\pi$-Berry phase is equivalent to a half magnetic flux-quanta threading through the 2D sample and encircled by the edge. According to the study by Wilczek~\cite{Wilczek}, a half quantum flux can lead to $1/2$-fractionalization of the orbital angular momentum. As a result, for the 2D sample with circular geometry, the orbital angular momentum of the edge modes should be fractionalized as $l_z=m+{\cal N}_{1d}/2$, with $m$ being integers. The fractionalization of the orbital angular momentum has an observable in the edge spectrum ${\cal E}_{l_z}=v_{\rm edge}l_zR_{l_z}^{-1}$, with $R_{l_z}$ the effective radius of edge state wave function. Due to the $1/2$-fractionalization no zero-energy (mid-gap) edge state exists, and therefore the total number of edge states is even. However, threading an additional magnetic $1/2$-flux-quanta can exactly push one original state to zero energy, changing the total number of edge modes to be odd, which provides an observable for the $1/2$-fractionalization of orbital angular momentum.

\begin{figure}[t]
\includegraphics[width=1\columnwidth]{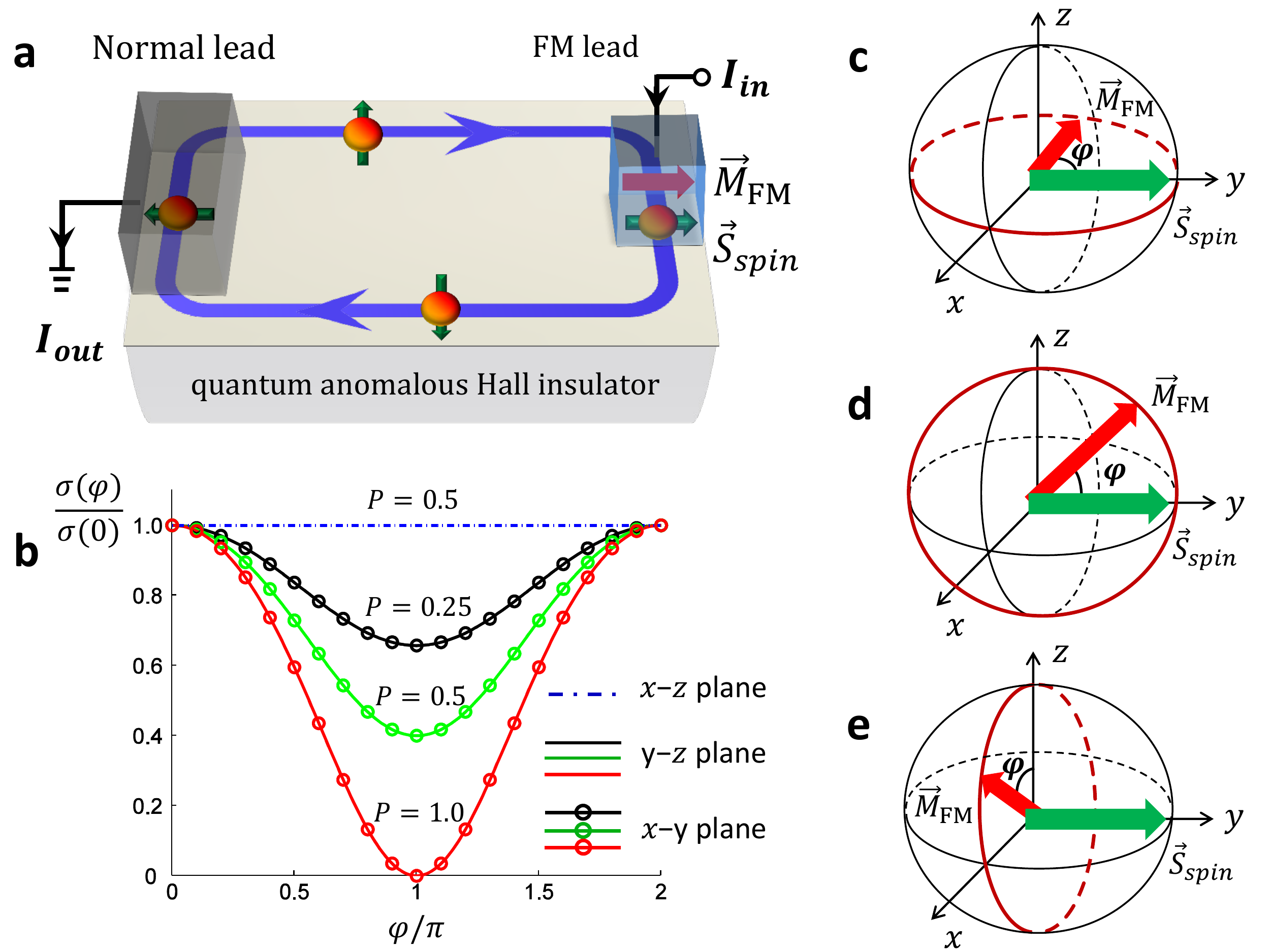}\caption{{(a)} A normal metallic lead is strongly coupled to the left-hand edge and a FM lead is weakly coupled to the right-hand edge of the QAH insulator. Here the parameters for the quantum anomalous Hall phase satisfy $m_z>0,A_{1,2}>0$, and $B<0$. Then spin of the edge states in the right-hand edge points to the $+y$ direction. {(b)} The tunneling conductance $\sigma(\varphi)$ is plotted versus azimuthal angle $\varphi$ of the magnetization in the FM lead, with the magnetization varying in the $x-y$ plane (c), $y-z$ plane (d), and $x-z$ plane (e), respectively. The numerical results are presented for different polarization ratios $P$ in the FM lead. The maximum tunneling conductance is obtained when the magnetization aligns with the edge spin-polarization direction.}
\label{Detection}
\end{figure}
The spin and orbital chirality make the edge of the QAH insulator be an exotic 1D metal which has no correspondence in conventional 1D materials. The topological spin texture of the edge modes may lead to strong spin-dependent effects as presented below, which on one hand can provide new unambiguous verification of the QAH state in the experiment, on the other hand, are applicable to spintronics by designing topological spin devices~\cite{Sankar2004}. As illustrated in Fig.~\ref{Detection} (a), we attach two metallic leads to the QAH sample, with a normal lead strongly coupled to the left-hand edge and a FM lead weakly coupled to the right-hand edge. Due to the spin texture, the couplings between the sample edge and leads are fully spin selective, which can lead to strong anisotropic effects in the tunneling conductance when changing the direction of magnetization $\vec M_{\rm FM}$ in the FM lead. The tunneling transport is studied with Landauer formalism with a square lattice tight-binding model whose low-energy limit gives Eq.~\eqref{eqn:Hamiltonian1}. With the coupling to normal and FM leads, we determine the retarded Green's function of the QAH insulator by
$G^R(\omega)=(\omega-H-\Sigma^R)^{-1}$, where $H$ is the tight binding Hamiltonian of the QAH insulator~\cite{SI}, and $\Sigma^R$ is the self-energy due to the couplings to the leads. Using the Fisher-Lee relation we can obtain the scattering matrix based on the Green's function and self-energies~\cite{lee1981}
\begin{equation}
S_{p,q}^{ss'}=-\delta_{p,q} \delta_{\alpha\beta }+i[ \Gamma_{p}^{s} ]^{1/2}G^R_{ss'}[ \Gamma_{q}^{s'}]^{1/2}.
\end{equation}
Here, the matrix element $S_{p,q}^{ss'}$ ($s,s'=\uparrow,\downarrow$) denotes the scattering amplitude of the process that an electron is scattered from the spin state $s'$ in lead $q$ to the spin state $s$ in lead $p$, with $p\neq q=L,R$ representing the left and right-hand leads, respectively. $\Gamma_{p}^{s}=i[(\Sigma_{p}^{ss})^R-(\Sigma_{p}^{ss})^A]$, where $(\Sigma_{p}^{ss})^{R/A}$ is the $s$-spin component retarded/advanced self-energy due to the coupling to lead $p$. From the scattering matrix we obtain the tunneling conductance by $\sigma(\varphi)=(e^2/h)\sum_{s,s'}|S_{p,q}^{ss'}|^2$, where $\varphi$ represents the direction of $\vec M_{\rm FM}$ in the FM lead.

\begin{figure}[h]
\includegraphics[width=1\columnwidth]{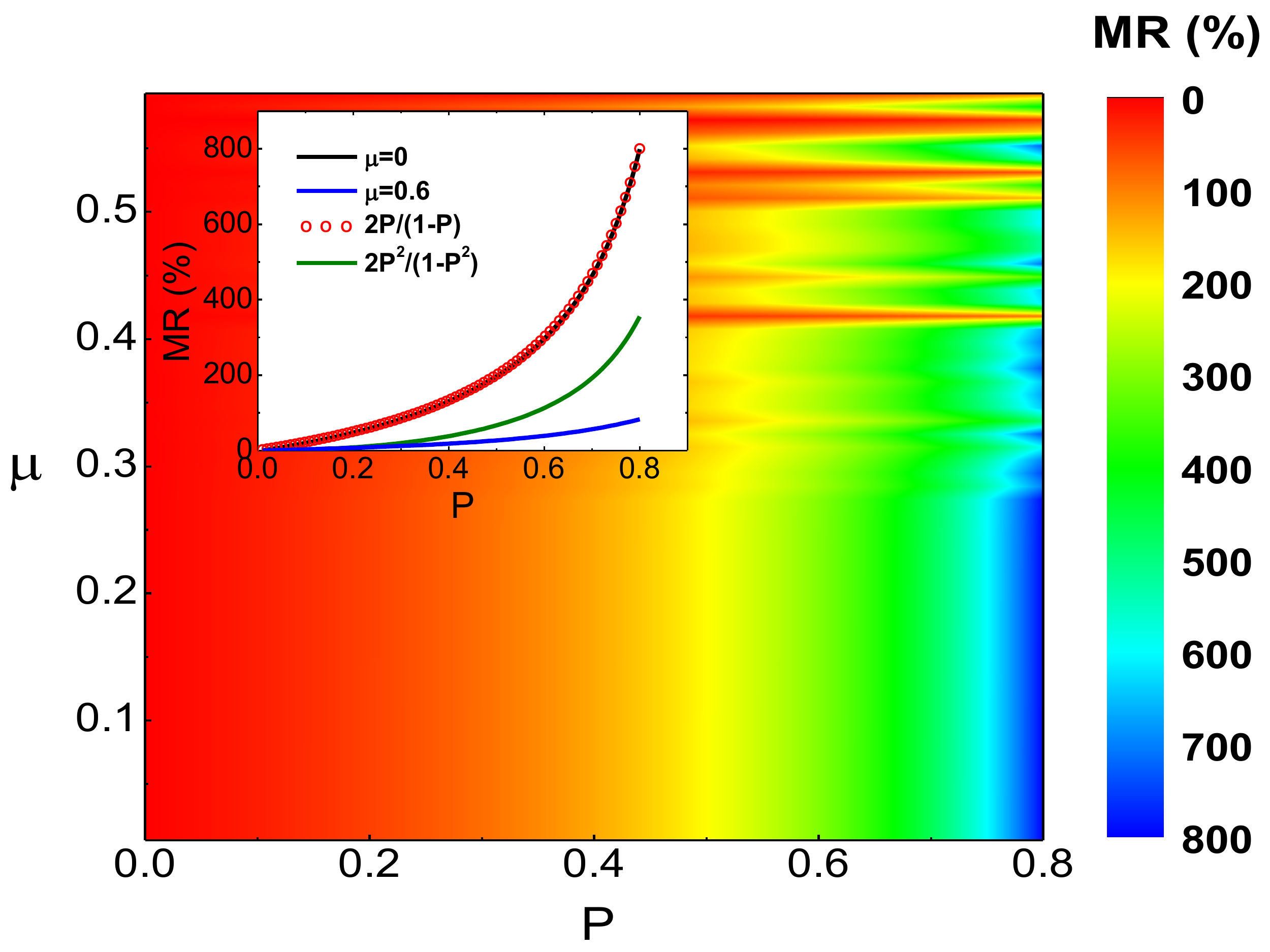}\caption{ Magnetoresistance for the setup in Fig.~\ref{Detection} (a) by setting magnetization of the FM lead along $\pm y$ directions. The magnetoresistance (MR) is plotted numerically as a function of the polarization ratio $P$ in the FM lead and the chemical potential $\mu$ (in unit of $B$) in the QAH insulator. The parameters for the QAH phase are taken as $m_z=-0.3A_{1,2}=-0.3B$, which gives the bulk gap $E_g=2|m_z|=0.6B$. The MR is uniform versus $\mu$ when the chemical potential is within the bulk gap ($|\mu|<0.3B$) and decreases when $\mu$ lies out of the gap. The inserted panel shows that the MR coincides with $2P/(1-P)$ for $|\mu|<0.3B$ (black solid and red circled curves), which is significantly larger than the corresponding tunneling MR, given by $2P^2/(1-P^2)$ (green curve), in the conventional ferromagnet/insulator/ferromagnet devices with polarization ratio $P$ in both ferromagnets~\cite{Julliere}.}
\label{magnetoresistance}
\end{figure}

The tunneling conductance $\sigma(\varphi)$ [Fig.~\ref{Detection} (b)] exhibits a clear angle-dependence when $\vec M_{\rm FM}$ varies in $x-y$ and $y-z$ planes [(c) and (d)], while it is a constant when $\vec M_{\rm FM}$ varies along $x-z$ plane (e). This measures that the edge spin polarizes to the $y$ direction. The angle-dependence implies a strong magnetoresistive effect by setting $\vec M_{\rm FM}$ along $\pm y$ directions. The magnetoresistance, given by ${\rm MR}=[\sigma(0)-\sigma(\pi)]/\sigma(\pi)\times100\%$, is plotted in Fig.~\ref{magnetoresistance} as a function of chemical potential $\mu$ in the QAH insulator and the polarization ratio $P$ of the FM lead. Due to the full spin-polarization, the edge of the sample can be regarded as an ideal dissipationless half-metal. This gives that ${\rm MR}\approx2P/(1-P)\times100\%$, which is significantly larger than the corresponding tunneling magnetoresistance obtained in conventional ferromagnet/insulator/ferromagnet devices with the same polarization ratio (the inserted panel of Fig.~\ref{magnetoresistance})~\cite{Julliere}. The strong magnetoresistive effect has been widely applied to spintronics, especially to designing read heads~\cite{Sankar2004}.

\begin{figure}[h]
\includegraphics[width=0.8\columnwidth]{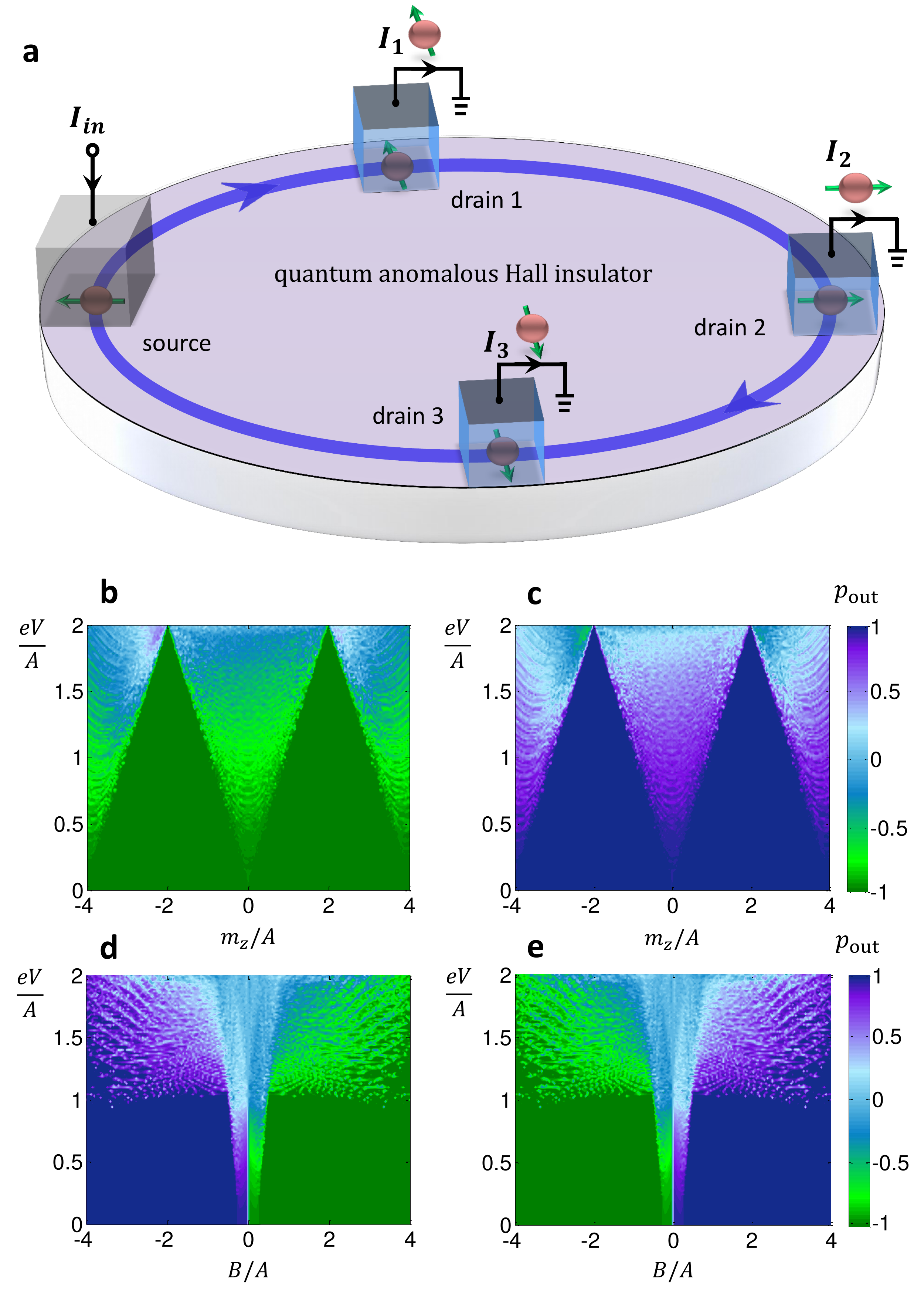}\caption{{Spin filtering effect and output spin-polarized current.} (a) For a quantum anomalous Hall insulator with circular boundary, the edge spin-polarization depends on the direction of the 1D edge. This provides a controllable way to generate spin-polarized current by attaching normal-metal leads to different directions of the sample edge. (b)-(e) The polarization ratio $p_{\rm out}$ of the output spin current is plotted as a function of voltage $eV$ in a drain lead and the Zeeman term $m_z$ (b,c) or $B$ (d,e), with the magnitudes rescaled by the SO coefficient $A$ ($=|A_{1,2}|$). Other parameters are taken as $B=A_1=A_2$ (a); $B=-A_1=-A_2$ (b); $m_z=A_1=A_2$ (c); and $m_z=-A_1=-A_2$ (d). The sign change of $p_{\rm out}$ in (d) and (e) from the topological region with $B<0$ to the region $B>0$ implies that the edge spin reverses direction.}
\label{spinfilter}
\end{figure}

Another interesting application of the topological spin texture is to design controllable spin-filtering devices, as illustrated in Fig.~\ref{spinfilter}. The edge spin texture ensures that the output current is fully spin-polarized, with the polarization direction depending on which edge the drain lead is attached to. The spin-polarization ratio of the output current can be calculated by $p_{\rm out}=\sum_{s}(|S_{p,q}^{s\uparrow}|^2-|S_{p,q}^{s\downarrow}|^2)/\sum_{s,s'}|S_{p,q}^{ss'}|^2$.
From the numerical results we see that when the voltage of the output lead lies in the sample bulk gap, the output current is $100\%$ polarized to the same direction, reflecting that the spin texture is identical for all edge states. We note that the band gap of the currently realized QAH effect is small, while for realistic applications a larger topological gap is necessary. The new physics unveiled in this work and the proposed potential applications will motivate the search for new novel materials of QAH insulator with sizable band gaps in the future.

In conclusion, we have predicted in the SO coupled QAH insulators that the edge states can exhibit topological spin textures in the boundary when a chiral-like symmetry is present, and found that such topological spin textures have a correspondence to the bulk topological phases. We also studied the tunneling transport from normal metal and FM leads to the chiral edge states and showed that the topological spin texture can induce strong magnetoresistance and spin filtering effects. These results may have potential applications in designing topological-state-based spin devices which might be applicable to future spintronic technologies.

We thank K. T. Law for helpful comments. XJL also appreciates the valuable discussions with Yayu Wang, Cenke Xu, Patrick A. Lee, Jairo Sinova, Zheng-Xin Liu, Meng Cheng, and X.-G. Wen. The authors thank the support of HKRGC through Grant 605512 and through Grant HKUST3/CRF/13G.

\noindent

\onecolumngrid

\renewcommand{\thesection}{S-\arabic{section}}
\setcounter{section}{0}  
\renewcommand{\theequation}{S\arabic{equation}}
\setcounter{equation}{0}  
\renewcommand{\thefigure}{S\arabic{figure}}
\setcounter{figure}{0}  

\indent

\section*{\Large\bf Supplementary Information}

\section{Two-band model for quantum anomalous Hall insulator}

The minimal realization for the quantum anomalous Hall effect is to consider a two-band model with spin-orbit coupling and magnetization. The main results in this work are not lattice configuration dependent. In this work we consider a square lattice tight-binding model with the Hamiltonian given by
\begin{eqnarray}\label{eqn:SItightbinding1}
H_{\rm QAH}&=&-B\sum_{<\bar{i},\vec{j}>}(\hat c_{\vec{i}\uparrow}^{\dag}\hat
c_{\vec{j}\uparrow}-\hat c_{\vec{i}\downarrow}^{\dag}\hat
c_{\vec{j}\downarrow})+\sum_{\vec{i}}(m_z+4B)(\hat n_{\vec{i}\uparrow}-\hat n_{\vec{i}\downarrow})+A_2\sum_{j_x}\bigr[(\hat c_{j_x\uparrow}^\dag\hat c_{j_x+1\downarrow}-\hat c_{j_x\uparrow}^\dag\hat c_{j_x-1\downarrow})+{\rm H.c.}\bigr]+\nonumber\\
&&+A_1\sum_{j_y}\bigr[i(\hat c_{j_y\uparrow}^\dag\hat c_{j_y+1\downarrow}-\hat c_{j_y\uparrow}^\dag\hat c_{j_y-1\downarrow})+{\rm H.c.}\bigr].
\end{eqnarray}
In the momentum space it reads $H=\sum_{\bold k}\psi_{\bold k}^\dag{\cal H}(\bold k)\psi_{\bold k}$,
where the spin basis $\psi_{\bold k}=(c_{\bold k\uparrow},c_{\bold k\downarrow})^T$ with $c_{\bold k,s}$ the electron annihilation operator in spin state $s=\uparrow, \downarrow$, and the Bloch Hamiltonian ${\cal H}(\bold k)$ takes the form ${\cal H}(\bold k)=2A_1\sin k_y\sigma_x-2A_2\sin k_x\sigma_y+\bigr[m_z+2B(2-\cos k_x-\cos k_y)\bigr]\sigma_z$.
Here we have taken the lattice constant to be $a=1$.
In the case $|m_z|\ll4|B|$, the physics of the system can be governed by the low-energy Hamiltonian
\begin{eqnarray}\label{eqn:SIHamiltonian1}
{\cal H}(\bold k)=2A_1k_y\sigma_x-2A_2k_x\sigma_y+\bigr[m_z+2B(k_x^2+k_y^2)\bigr]\sigma_z.
\end{eqnarray}
The Bloch Hamiltonian can be rewritten in the form ${\cal H}(\bold k)=\vec d(\bold k)\cdot\vec\sigma$, where the $\vec d$-vector is defined through $d_x=2A_1 k_y, d_y=-2A_2k_x$, and $d_z=m_z+2B(k_x^2+k_y^2)$. The topology of the system is determined by the first Chern number, which can be calculated by $C_1=\frac{1}{4\pi}\int d^2\bold k\bold n\cdot\frac{\partial\bold n}{\partial{k_x}}\times\frac{\partial\bold n}{\partial{k_y}},  \ \bold n=\frac{1}{|\vec d(\bold k)|}(d_x,d_y,d_z)$.
The integrand in the right hand side of the above equation describes a mapping between the momentum $\bold k$-space to the spherical surface $S^2$ formed by the unit vector $\bold n(\bold k)$. Therefore the first Chern invariant is given by the number of times that this mapping can cover the whole spherical surface. For each fixed momentum $k=(k_x^2+k_y^2)^{1/2}$, the $\bold n$-vector can cover one circle in the $x-y$ plane. Moreover, it is easy to see that for both $k=0$ and $k=\infty$, the unit vector $\bold n(\bold k)$ points along the $z$ or $-z$ axis. Therefore the mapping must cover integer times of the whole spherical surface. In particular, when $m_zB>0$, the $z$-component of the $\bold n$-vector is always positive (for $m_z>0$) or negative (for $m_z<0$), and the number of coverage is zero. This gives that $C_1=0$. On the other hand, for $m_zB<0$ and $m_z>0$, the unit vector $\bold n$ points to $+z$ and $-z$ directions when $k=0$ and $k=\infty$, respectively. We then get a single coverage of the spherical surface in the mapping, and obtain the Chern number $C_1=1$. Finally, it can be seen that the Chern number is odd with respect to each component of $\bold n(\bold k)$, and it changes sign if reversing the sign of any component $n_j$. With these results in mind we conclude that
\begin{eqnarray}\label{eqn:conductivity}
C_1=\mbox{sgn}(A_1A_2)[\mbox{sgn}(m_z)-\mbox{sgn}(B)]/2.
\end{eqnarray}
The Hall conductance of the quantum anomalous Hall insulator is given by $\sigma_{xy}=C_1e^2/h$, where $e$ and $h$ are the electron charge and Plank constant, respectively.

\section{Topological edge spin texture}

\begin{figure}[h]
\includegraphics[width=0.8\columnwidth]{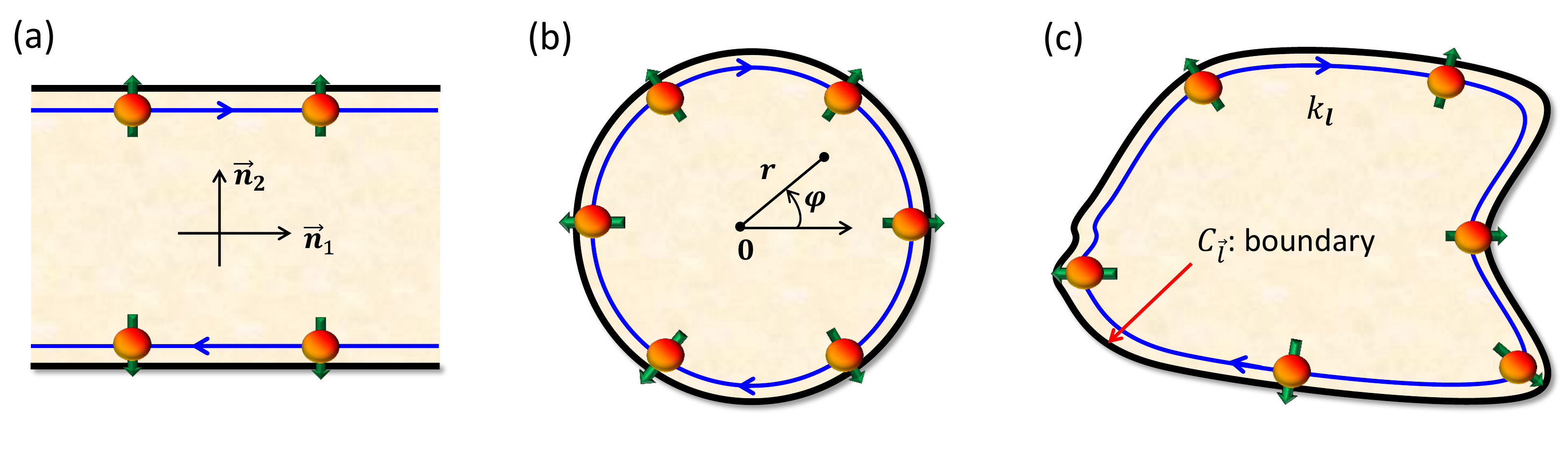}\caption{Edge spin polarizations for the quantum anomalous Hall insulator with different boundary geometries. (a) The open boundary which has two edges normal to $\vec n_2$ direction, and is infinite (or periodic) along $\vec n_1$ axis; (b) The circular closed boundary which is rotationally invariant; (c) The generic boundary. In the illustration we consider the parameter regime that $m_z>0, B<0$, and $A_{1,2}>0$.}
\label{boundaries}
\end{figure}

We consider the isotropic low-energy Dirac Hamiltonian~\eqref{eqn:SIHamiltonian1} with $A_1=\eta A_2$, where $\eta=\pm1$. In this case the low-energy Hamiltonian is rotationally invariant and can be written down in the following generic form
\begin{eqnarray}\label{eqn:SIHamiltonian2}
{\cal H}(\bold k)=2A_{1}k_{n_2}\sigma_{n_1}-2A_{2}k_{n_1}\sigma_{n_2}+(m_z+2Bk^2)\sigma_z,
\end{eqnarray}
where the {\it arbitrary} in-plane orthogonal unit vectors are $\vec n_1=u n_x+v n_y$ and $\vec n_2=-v n_x+u n_y$, where the coefficients satisfy $u^2+v^2=1$. The components of the momenta and Pauli matrices are given by $\sigma_{n_1}=\vec n_1\cdot\vec\sigma, \sigma_{n_2}=\vec n_2\cdot\vec\sigma,
k_{n_1}=uk_x+v\eta k_y$, and $k_{n_2}=-v\eta k_x+uk_y$.
It can be verified that the Hamiltonian~\eqref{eqn:SIHamiltonian2} respects the chiral-like symmetry defined by $S=\sigma_{n_2}{\cal M}_{n_1}$ (also valid for $\sigma_{n_1}{\cal M}_{n_2}$), with ${\cal M}_{n_1}$ representing the spatial reflection with respect to the $\vec n_1$ direction. It follows that
\begin{eqnarray}\label{eqn:SIsymmetry1}
S{\cal H}(\bold k) S^\dag=-{\cal H}(\bold k).
\end{eqnarray}
Note that in the second quantization picture the symmetry transforms via $S(c_{\bold ks},c^\dag_{\bold ks})^T=(\sigma_{\vec n_2})_{ss'}(c^\dag_{\bold k's'},c_{\bold k's'})^T$, with $\bold k'=(-k_{n_1},k_{n_2})$, with which one can verify for the second quantization Hamiltonian that $SHS^{\dag}=H$.
We show below that due to this symmetry the edge states of the quantum anomalous Hall insulator are in-plane spin-polarized, and exhibit topological spin texture in the boundary. In particular, we consider three different situations as illustrated in Fig.~\ref{boundaries}: the open boundary (a), the circular close boundary (b), and the generic closed boundary (c). Furthermore, we should emphasize that while we use here the Hamiltonian~\eqref{eqn:SIHamiltonian2} as an example for the study, which is relevant for the realistic experiment. the edge spin polarizations shown below are generic results for quantum anomalous Hall insulators satisfying the chiral-like symmetry given in Eq.~\eqref{eqn:SIsymmetry1}.

\subsection{Open boundary}

We first consider the simplest situation that the system has two edges normal to $\vec n_2$ direction, while it is infinite (or periodic) along $\vec n_1$ axis [Fig.~\ref{boundaries} (a)]. This study can be applied to the system with smooth and slowly varying (closed) boundaries. In this case the momenta $k_{n_1}$ are good quantum numbers, and the Hamiltonian can be described by
\begin{eqnarray}\label{eqn:SIsymmetry2}
H=\sum_{k_{n_1}}H_{1D}(k_{n_1}),
\end{eqnarray}
where $H_{1D}(k_{n_1})$ is a $k_{n_1}$-parameterized 1D Hamiltonian. The end states of $H_{1D}(k_{n_1})$ are edge states of the original 2D quantum anomalous Hall system with momentum $k_{n_1}$. The 1D Hamiltonian can be expressed as
\begin{eqnarray}\label{eqn:SIHamiltonian1D1}
H_{1D}(k_{n_1})&=&-\int{dx_{n_2}}\biggr\{\psi_{k_{n_1}}^{\dag}(x_{n_2})\bigr(2B\partial^2_{x_{n_2}}\sigma_z+iA_1\partial_{x_{n_2}}\sigma_{n_1}\bigr)\psi_{k_{n_1}}(x_{n_2})\biggr\}\nonumber\\
&&+\int{dx_{n_2}}\biggr\{\psi_{k_{n_1}}^{\dag}(x_{n_2})\bigr[(m_z+2Bk_{n_1}^2)\sigma_z-A_2k_{n_1}\sigma_{n_2}\bigr]\psi_{k_{n_1}}(x_{n_2})\biggr\}.
\end{eqnarray}
The transformation in Eq.~\eqref{eqn:SIsymmetry1} is defined for the first quantization Hamiltonian. Accordingly, for the second quantization Hamiltonian, the symmetry operator $S$ transforms the basis according to $S(c_{k_{n_1}\uparrow},c^\dag_{k_{n_1}\downarrow})^T=(c^\dag_{-k_{n_1}\downarrow},c_{-k_{n_1}\uparrow})^T$. This is followed by
\begin{eqnarray}\label{eqn:SIsymmetry3}
SH_{1D}(k_{n_1})S^\dag=H_{1D}(-k_{n_1}),
\end{eqnarray}
which gives that $SHS^\dag=H$.
It is trivial to know that at $k_{n_1}=0$, which is a reflection invariant momentum, the 1D Hamiltonian $H_{1D}(0)$ is invariant under the $S$-transformation. This implies that the system $H_{1D}(0)$ belongs to the 1D chiral unitary (AIII) class~\cite{Schnyder,Liu2013AIII}. Indeed, this can be more transparent if rotating the spin operators $(\sigma_{n_1},\sigma_{n_2})\rightarrow(\sigma_y,-\sigma_x)$ in $H_{1D}(0)$. In this case, one can verify that the time-reversal and charge conjugation symmetries, defined respectively by ${\cal T}=iK\sigma_y$ with $K$ the
complex conjugation, and ${\cal C}: (\hat c_{k_{n_1}\sigma},\hat c^\dag_{k_{n_1}\sigma})\rightarrow(\sigma_z)_{\sigma\sigma'}(\hat c^\dag_{-k_{n_1}\sigma'},\hat c_{-k_{n_1}\sigma'})$, are generically broken for $H_{1D}$, while at $k_{n_1}=0$ the chiral symmetry $S={\cal T}{\cal C}$ is preserved. Therefore $H_{1D}(0)$ describes 1D topological insulating phase belonging to the chiral unitary (AIII) class, whose end states are eigenstates of the chiral operator $\sigma_{n_2}$, with the spin oppositely polarized in the opposite edges~\cite{Liu2013AIII}.

The spin-polarization of the edge states with $k_{n_1}\neq0$ can be obtained using the $\bold k\cdot\bold p$ theory. By expanding up to the leading order of the momentum $k_{n_1}$ the effective edge Hamiltonian for the two edges normal to $\vec n_2$ direction, under the restriction of the $S$ symmetry, should take the generic form $v_{\rm edge}k_{\vec n_1}\sigma_{\vec n_2}$. Therefore the edge states with nonzero $k_{\vec n_1}$ are also eigenstates of $\sigma_{\vec n_2}$.
The edge spin polarizations are illustrated in Fig.~\ref{boundaries} (a) under the condition with $m_z>0, B<0$, and $A_{1,2}>0$. We note that the Eqs.~\eqref{eqn:SIsymmetry2} and~\eqref{eqn:SIsymmetry3} are directly derived from the symmetry given in Eq.~\eqref{eqn:SIsymmetry1}, not dependent on the specific Hamiltonian~\eqref{eqn:SIHamiltonian2} or~\eqref{eqn:SIHamiltonian1D1}. Therefore the edge spin polarizations are generic results for any 2D Hamiltonian satisfying the $S$ symmetry.

\subsection{Circular closed boundary}

We turn to the edge spin texture for the finite system with closed boundary. For convenience we study the quantum anomalous Hall sample with circular geometry, and show below that the edge modes exhibit topological spin texture as illustrated in Fig.~\ref{boundaries} (b). Noting that the boundary is rotationally invariant, it is convenient to reexpress the Hamiltonian in the polar coordinate $(r,\varphi)$ system
\begin{eqnarray}\label{eqn:SIHamiltonian3}
{\cal H}(r,\varphi)=\bigr[2B(\frac{1}{r}\partial_rr\partial_r+\frac{1}{r^2}\partial^2_{\varphi})+m_z\bigr]\sigma_z+i\frac{2A_{1}}{r}\sigma_{r}\partial_\varphi -i2A_{2}\sigma_{\varphi}\partial_{r},
\end{eqnarray}
where $\sigma_r=\cos\varphi\sigma_x+\sin\varphi\sigma_y$ and $\sigma_\varphi=\cos\varphi\sigma_y-\sin\varphi\sigma_x$. The eigenstates of ${\cal H}(r,\varphi)$ can generically be described by
\begin{eqnarray}\label{eqn:SIeigenstate1}
|\Phi_{m}(r,\varphi)\rangle=|\phi_{m}(r,\varphi)\rangle e^{im\varphi},
\end{eqnarray}
where $m$ are integers. For bulk states, each $m$ corresponds to two solutions $|\phi^{(\pm)}_{m}(r,\varphi)\rangle$, with eigenvalues $\pm E_m$ ($E_m>0$) respectively. On the other hand, for edge modes each $m$ corresponds to only a single eigenstate denoted by $|\phi^{\rm edge}_{m}(r,\varphi)\rangle$. These states can be solved by
\begin{eqnarray}\label{eqn:SIeigenstate2}
\biggr\{\bigr[2B(\frac{1}{r}\partial_rr\partial_r-\frac{m^2-i2m\partial_{\varphi}-\partial_{\varphi}^2}{r^2})+m_z\bigr]\sigma_z -i2A_{2}\sigma_{\varphi}\partial_{r}-\frac{2A_{1}}{r}\sigma_{r}(m-i\partial_\varphi)\biggr\}|\phi^{(\pm)}_{m}(r,\varphi)\rangle=\pm E_m|\phi^{(\pm)}_{m}(r,\varphi)\rangle,
\end{eqnarray}
and
\begin{eqnarray}\label{eqn:SIeigenstate3}
\biggr\{\bigr[2B(\frac{1}{r}\partial_rr\partial_r-\frac{m^2-i2m\partial_{\varphi}-\partial_{\varphi}^2}{r^2})+m_z\bigr]\sigma_z -i2A_{2}\sigma_{\varphi}\partial_{r}-\frac{2A_{1}}{r}\sigma_{r}(m-i\partial_\varphi)\biggr\}|\phi^{\rm edge}_{m}(r,\varphi)\rangle={\cal E}_m|\phi^{\rm edge}_{m}(r,\varphi)\rangle.
\end{eqnarray}
For the present circular boundary, the chiral-like symmetry is given by $S=\sigma_r{\cal M}_\varphi$, where ${\cal M}_\varphi$ transforms $\varphi$ to $-\varphi$. Using $S$ to act on both sides of the eigen-equations for the bulk and edge states yields
\begin{eqnarray}\label{eqn:SIeigenstate2'}
\biggr\{\bigr[2B(\frac{1}{r}\partial_rr\partial_r-\frac{m^2+i2m\partial_{\varphi}-\partial_{\varphi}^2}{r^2})+m_z\bigr]\sigma_z -i2A_{2}\sigma_{\varphi}\partial_{r}+\frac{2A_{1}}{r}\sigma_{r}(m+i\partial_\varphi)\biggr\}|\phi^{(\pm)}_{-m}(r,-\varphi)\rangle=\pm E_m|\phi^{(\pm)}_{-m}(r,-\varphi)\rangle,\nonumber\\
\end{eqnarray}
and
\begin{eqnarray}\label{eqn:SIeigenstate3'}
\biggr\{\bigr[2B(\frac{1}{r}\partial_rr\partial_r-\frac{m^2+i2m\partial_{\varphi}-\partial_{\varphi}^2}{r^2})+m_z\bigr]\sigma_z -i2A_{2}\sigma_{\varphi}\partial_{r}+\frac{2A_{1}}{r}\sigma_{r}(m+i\partial_\varphi)\biggr\}|\phi^{\rm edge}_{-m}(r,-\varphi)\rangle=-{\cal E}_m|\phi^{\rm edge}_{-m}(r,-\varphi)\rangle.\nonumber\\
\end{eqnarray}
Taking that $\varphi\rightarrow-\varphi$, we can rewrite the above equations in the form
\begin{eqnarray}\label{eqn:SIeigenstate4}
\biggr\{\bigr[2B(\frac{1}{r}\partial_rr\partial_r-\frac{m^2-i2m\partial_{\varphi}-\partial_{\varphi}^2}{r^2})+m_z\bigr]\sigma_z -i2A_{2}\sigma_{\varphi}\partial_{r}+\frac{2A_{1}}{r}\sigma_{r}(m-i\partial_\varphi)\biggr\}|\phi^{(\pm)}_{-m}(r,\varphi)\rangle=\pm E_m|\phi^{(\pm)}_{-m}(r,\varphi)\rangle,
\end{eqnarray}
and
\begin{eqnarray}\label{eqn:SIeigenstate5}
\biggr\{\bigr[2B(\frac{1}{r}\partial_rr\partial_r-\frac{m^2-i2m\partial_{\varphi}-\partial_{\varphi}^2}{r^2})+m_z\bigr]\sigma_z -i2A_{2}\sigma_{\varphi}\partial_{r}+\frac{2A_{1}}{r}\sigma_{r}(m-i\partial_\varphi)\biggr\}|\phi^{\rm edge}_{-m}(r,\varphi)\rangle=-{\cal E}_m|\phi^{\rm edge}_{-m}(r,\varphi)\rangle.
\end{eqnarray}
From the last four formulas we find that the bulk states $|\phi^{(+)}_{m}(r,\varphi)$ and $|\phi^{(-)}_{-m}(r,-\varphi)$ have opposite energies $E_m$ and $-E_m$, respectively. For edge states we have opposite energies ${\cal E}_m$ and $-{\cal E}_m$ for $|\phi^{\rm edge}_{m}(r,\varphi)\rangle$ and $|\phi^{\rm edge}_{-m}(r,-\varphi)\rangle$, respectively. On the other hand, generically we have $E_m\neq E_{-m}$ and ${\cal E}_m\neq-{\cal E}_{-m}$ (in Eqs.~\eqref{eqn:SIeigenstate4} and~\eqref{eqn:SIeigenstate5} the left-hand side is not the original Hamiltonian). This implies that for the present circular boundary, in the edge state spectrum ${\cal E}_0\neq0$. From Eqs.~\eqref{eqn:SIeigenstate2'} and~\eqref{eqn:SIeigenstate3'} we have that the corresponding eigenstates transform according to
\begin{eqnarray}\label{eqn:SItransform1}
S|\phi^{(\pm)}_{m}(r,\varphi)\rangle=|\phi^{(\mp)}_{-m}(r,-\varphi)\rangle, \ S|\phi^{\rm edge}_{m}(r,\varphi)\rangle=|\phi^{\rm edge}_{-m}(r,-\varphi)\rangle.
\end{eqnarray}
Furthermore, we note that the reflection operator ${\cal M}_\varphi$ transforms the bulk and edge states via
\begin{eqnarray}\label{eqn:SItransform2}
{\cal M}_\varphi|\phi^{(\pm)}_{m}(r,\varphi)\rangle=|\phi^{(\pm)}_{-m}(r,-\varphi)\rangle, \ {\cal M}_\varphi|\phi^{\rm edge}_{m}(r,\varphi)\rangle=|\phi^{\rm edge}_{-m}(r,-\varphi)\rangle.
\end{eqnarray}
Together with the results in Eqs.~\eqref{eqn:SItransform1} and \eqref{eqn:SItransform2} we can deduce for the bulk and edge states that (up to $\pm$ signs)
\begin{eqnarray}
\sigma_r|\phi^{(\pm)}_{m}(r,\varphi)\rangle=|\phi^{(\mp)}_{m}(r,\varphi)\rangle, \ \sigma_r|\phi^{\rm edge}_{m}(r,\varphi)\rangle=|\phi^{\rm edge}_{m}(r,\varphi)\rangle.
\end{eqnarray}
This shows that the edge modes are eigenstates of $\sigma_r$ and then the edge spin polarization takes the spatial configuration illustrated in Fig.~\ref{boundaries} (b) (with the parameter regime that $m_z>0, B<0$, and $A_{1,2}>0$), while the bulk states are generically not eigenstates of $\sigma_r$ and do not exhibit topological spin texture in the position space.

\subsection{Generic boundary}

Now we generalize the above results to the situation with generic boundary geometries [Fig.~\ref{boundaries} (c)]. We require that in the sample there are no narrow areas where the edge modes localized in different edges may couple to each other, leading to the finite size effects. Let the boundary be characterized by a curve $C_{\vec l}: f(r,l)=0$, with $(r,l)$ consisting of a local orthogonal coordinate system. Around the boundary, one can always locally describe the Hamiltonian by
\begin{eqnarray}\label{eqn:SIHamiltonian4}
{\cal H}(r,l)=\bigr[2B(p_r^2+p_l^2)+m_z\bigr]\sigma_z+2A_{1}p_l\sigma_{r}-2A_{2}p_{r}\sigma_{l},
\end{eqnarray}
where $p_l$ and $p_r$ are momentum operators with respect to the local tangent ($\vec n_l$) and normal ($\vec n_r$) directions of the boundary, respectively. The Pauli matrices $\sigma_r=\vec n_r\cdot\vec\sigma$ and $\sigma_l=\vec n_l\cdot\vec\sigma$. For the above Hamiltonian the symmetry operator is defined by $S=\sigma_r{\cal M}_l$. Similarly, the edge states can be described by $|\Phi^{\rm edge}_{k_l}(r,l)\rangle$, with $k_l$ the quasi-momentum along the boundary. Through the same procedure as done in the above subsection, we can verify that the eigenvalues for $|\Phi^{\rm edge}_{k_l}(r,l)\rangle$ and $|\Phi^{\rm edge}_{-k_l}(r,-l)\rangle$ are ${\cal E}_{k_l}$ and $-{\cal E}_{k_l}$, respectively. The transformations on these states satisfy $S|\Phi^{\rm edge}_{k_l}(r,l)\rangle=|\Phi^{\rm edge}_{-k_l}(r,-l)\rangle$ and ${\cal M}_l|\Phi^{\rm edge}_{k_l}(r,l)\rangle=|\Phi^{\rm edge}_{-k_l}(r,-l)\rangle$. Therefore, we again have $\sigma_r|\Phi^{\rm edge}_{k_l}(r,l)\rangle=|\Phi^{\rm edge}_{k_l}(r,l)\rangle$ up to a $\pm$ sign. These results conclude that the edge states exhibit topological spin texture in the boundary.

The topological edge spin texture leads to a nontrivial quantized Berry phase in the boundary, which can be calculated by
\begin{eqnarray}\label{eqn:Berry}
\gamma_C=\oint d\tilde x{\cal A}_s(\tilde x), \ {\cal A}_s=i\hbar\langle\chi_s(\tilde x|\partial_x|\chi_s(\tilde x)\rangle,
\end{eqnarray}
where we have denoted by $|\chi_s(\tilde x)\rangle$ the spin part of the edge state wave function, with $\tilde x$ the coordinate along the boundary. Due to the topological spin texture, the Berry phase reads $\gamma_C=\pm\pi$, which defines a 1D topological state characterized by the winding number ${\cal N}_{1d}=\gamma_C/\pi$ in the edge. Note that the Berry phase is determined by the two in-plane spin components, it should be odd under the transformation by ${\cal M}_{n_2}$ or ${\cal M}_{n_1}$, and be even under the transformation $\sigma_z\rightarrow-\sigma_z$. Therefore the 1D winding number is given by ${\cal N}_{1d}=\mbox{sgn}(A_1A_2)$, and it has a correspondence to the bulk Chern number via
\begin{eqnarray}\label{eqn:Correspondence}
C_1=\mbox{sgn}(m_z){\cal N}_{1d}.
\end{eqnarray}
It is worthwhile to note that there is an additional sign factor in the correspondence between the 2D bulk topological state and the 1D topological state in the boundary. This is reasonable, since the first Chern invariant is a 2D winding number, and it necessitates the inclusion of one more dimension relative to that for ${\cal N}_{1d}$. The additional sign factor accounts for such difference in the dimension.

\section{Symmetry-breaking perturbations}

In the low-energy regime, the nonperturbed Hamiltonian is rotationally invariant and respect the chiral-like $S$ symmetry. The local perturbation which breaks this symmetry includes the in-plane Zeeman fields, the nonmagnetic and magnetic disorders. In the high-energy regime, due to discrete lattice anisotropy the chiral-like symmetry is generically broken in the Bloch Hamiltonian, unless for several special directions.

\subsection{In-plane magnetic fields}

The in-plane Zeeman couplings $V_{\rm pert}=m_x\sigma_x+m_y\sigma_y$ can break the $S$ symmetry. For the sake of generality, we write down the Hamiltonian in the bases of momenta and Pauli matrices with respect to the generic orthogonal unit vectors $\vec n_1$ and $\vec n_2$
\begin{eqnarray}
{\cal H}(\bold k)=(2A_{1}k_{n_2}+m_1)\sigma_{n_1}-(2A_{2}k_{n_1}+m_2)\sigma_{n_2}+(m_z+2Bk^2)\sigma_z,
\end{eqnarray}
where $m_1=um_x+vm_y$ and $m_2=vm_x-um_y$. The in-plane Zeeman fields shift the bulk band edge from the $\Gamma$ point to the one with finite momenta. In the parameter regime $m_zB<0$, increasing the Zeeman field can lead to a topological phase transition at the critical point
\begin{eqnarray}
m_\parallel=m_{\parallel}^c=2^{1/2}A\frac{|m_z|^{1/2}}{|B|^{1/2}},
\end{eqnarray}
where $m_\parallel=(m_x^2+m_y^2)^{1/2}$ and $A=|A_{1,2}|$. The condition with $m_\parallel<m_\parallel^c$ corresponds to the topological phase. Similar as the study in the subsection II (A), we consider the boundary modes localized in the two edges normal to $\vec n_2$ direction. It is straightforward to verify that for $S=\sigma_{\sigma_{n_2}}{\cal M}_{n_1}$ we have
\begin{eqnarray}
S{\cal H}(\bold k)S^\dag=-{\cal H}(\bold k)-2m_2\sigma_{n_2}.
\end{eqnarray}
Therefore, in the presence of in-plane Zeeman fields, generically no symmetry is preserved for the system. However, it is interesting that the Hamiltonian also satisfies
\begin{eqnarray}
\sigma_{n_2}{\cal H}(\bold k)\sigma_{n_2}^{-1}=-{\cal H}(\bold k)-2(m_2+2A_2k_{n_1})\sigma_{n_2}.
\end{eqnarray}
This implies that the Hamiltonian ${\cal H}_{1D}(k^0_{n_1},x_{n_2})$ with $k^0_{n_1}=-m_2/(2A_2)$ again defines a 1D chiral unitary (AIII) class insulator~\cite{Schnyder,Liu2013AIII}. Then in the topological phase the edge states localized in the left-hand side $|\Phi^{\rm edge}_{k^0_{n_1}}(x_2)\rangle_L$ and right-hand side $|\Phi^{\rm edge}_{k^0_{n_1}}(x_2)\rangle_R$ are eigenstates of $\sigma_{n_2}$, with the spin oppositely polarized in the opposite edges~\cite{Liu2013AIII}. Furthermore, for the generic 1D momentum $k_{n_1}=k_{n_1}^0+\tilde k_{n_1}$, we shift the zero point of the momentum to $k_{n_1}^0$, and can then rewrite the Hamiltonian in the form
\begin{eqnarray}\label{eqn:SIHamiltonian1D2}
H_{1D}(\tilde k_{n_1},x_{n_2})&=&-\int{dx_{n_2}}\biggr\{\psi_{k_{n_1}}^{\dag}(x_{n_2})\bigr(2B\partial^2_{x_{n_2}}\sigma_z+iA_1\partial_{x_{n_2}}\sigma_{n_1}\bigr)\psi_{k_{n_1}}(x_{n_2})\biggr\}\nonumber\\
&&+\int{dx_{n_2}}\biggr\{\psi_{k_{n_1}}^{\dag}(x_{n_2})\bigr[(\tilde m_z+2B\tilde k_{n_1}^2)\sigma_z-A_2\tilde k_{n_1}\sigma_{n_2}\bigr]\psi_{k_{n_1}}(x_{n_2})\biggr\},
\end{eqnarray}
where $\tilde m_z=m_z+4Bk_{n_1}^0\tilde k_{n_1}+2B\tilde k_{n_1}^2$. It can be seen that the above 1D Hamiltonian is {\it formally} equivalent to the one in Eq.~\eqref{eqn:SIHamiltonian1D1}. Therefore, according to the previous results, all the boundary states localized in the edges normal to $\vec n_2$ direction are eigenstates of $\sigma_{n_2}$. With this result we confirm that the edge spin texture cannot be affected by the in-plane Zeeman fields without driving the phase transition in the bulk.

\subsection{Magnetic and non-magnetic disorders}

The nonmagnetic and magnetic disorders can also break the chiral-like symmetry $S$ of the system. For convenience we adopt the lattice model to study the disorder effects. The local on-site magnetic and nonmagnetic disorders can be described by
\begin{eqnarray}\label{eqn:SIdisorder1}
V_{\rm dis}=\sum_{\vec r_j,s}V^{\rm non}_{\rm dis}(\vec r_j)n_{\vec r_j,s}+\sum_{\vec r_j,\alpha}V^{\rm mag}_{\rm dis,\alpha}(\vec r_j)\psi_{\vec r_j}^\dag\sigma_\alpha\psi_{\vec r_j},
\end{eqnarray}
where $\psi_{\vec r_j}=(c_{\vec r_j,\uparrow},c_{\vec r_j,\downarrow})$, the particle number operator $n_{\vec r_j,s}=c_{\vec r_j,s}^\dag c_{\vec r_j,s}$ with $\vec r_j$ the 2D coordinate for lattice sites, $s=\uparrow,\downarrow,$ and $\alpha=x,y,z$. Here $V^{\rm non}_{\rm dis}(\vec r_j)\in[-V_0/2,V_0/2]$ and $V^{\rm mag}_{\rm dis,\alpha}(\vec r_j)\in[-V_\alpha/2,V_\alpha/2]$ represent nonmagnetic and magnetic random disorder potentials, respectively, and for magnetic disorder we take that $V_x^2+V_y^2+V_z^2=V_0^2$. The configuration averaging of the disorder potentials vanishes
\begin{eqnarray}
\langle V^{\rm non}_{\rm dis}(\vec r_j)\rangle=\langle V^{\rm mag}_{\rm dis,\alpha}(\vec r_j)\rangle=0.
\end{eqnarray}

In the regime that the disorder strength is weak relative to the bulk gap of the quantum anomalous Hall insulator, we expect that the chiral edge spin texture cannot be scattered. The reason is because the spin chirality of the edge states ensures that edge modes with opposite spin polarizations are spatially localized far away from each other, and weak disorder, while breaking the symmetry of the bulk Hamiltonian, cannot lead to the scattering between such two edge modes. If we consider only the nonmagnetic disorder, the commutation relation between $S=\sigma_{n_2}{\cal M}_{n_1}$ and the second quantization Hamiltonian satisfies
\begin{eqnarray}\label{eqn:SIdisorderrelation1}
SHS^\dag=H-\sum_{\vec r_j,s}V^{\rm non}_{\rm dis}(\vec r_j)n_{\vec r_j,s},
\end{eqnarray}
where the condition $\langle V^{\rm non}_{\rm dis}(\vec r_j)\rangle=0$ has been applied.
On the other hand, if there is only magnetic disorder, we have
\begin{eqnarray}\label{eqn:SIdisorderrelation2}
SHS^\dag=H-\sum_{\vec r_j}V^{\rm mag}_{\rm dis,n_2}(\vec r_j)\psi_{\vec r_j}^\dag\sigma_{n_2}\psi_{\vec r_j}.
\end{eqnarray}
Comparing with the formulas~\eqref{eqn:SIdisorderrelation1} and~\eqref{eqn:SIdisorderrelation2} we can see that under the same strength of disorder potentials, the topological spin texture should be more insensitive to the magnetic disorders. This is because, for example, for the edges normal to $\vec n_2$ direction, only the magnetic disorder component $V^{\rm mag}_{n_2}$ breaks the $S$ symmetry and may affect the edge spin polarizations. In particular, the magnetic disorder with polarization along $z$ direction has no effect on the edge spin texture.

\begin{figure}[h]
\includegraphics[width=1.0\columnwidth]{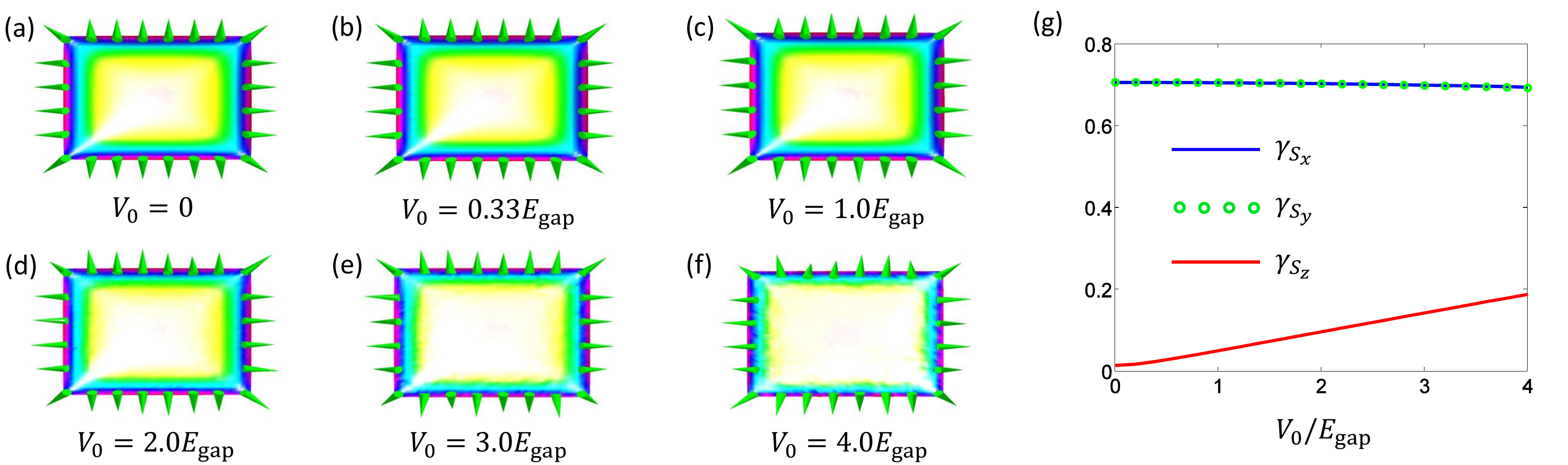}\caption{Effect of nonmagnetic disorder scattering on the edge spin texture. The parameters for the numerical calculation are rescaled to be dimensionless and are taken as $A_1=A_2=-B=1.0$ and $m_z=0.3$. (a-f) The edge spin configuration with different magnitudes of the nonmagnetic disorder potential $V_0$. The colors represent the wave-function distribution of the edge states; (g) Root Mean Square for $S_x, S_y$ and $S_z$ versus $V_0$.}
\label{nonmagdisorder}
\end{figure}
\begin{figure}[h]
\includegraphics[width=1.0\columnwidth]{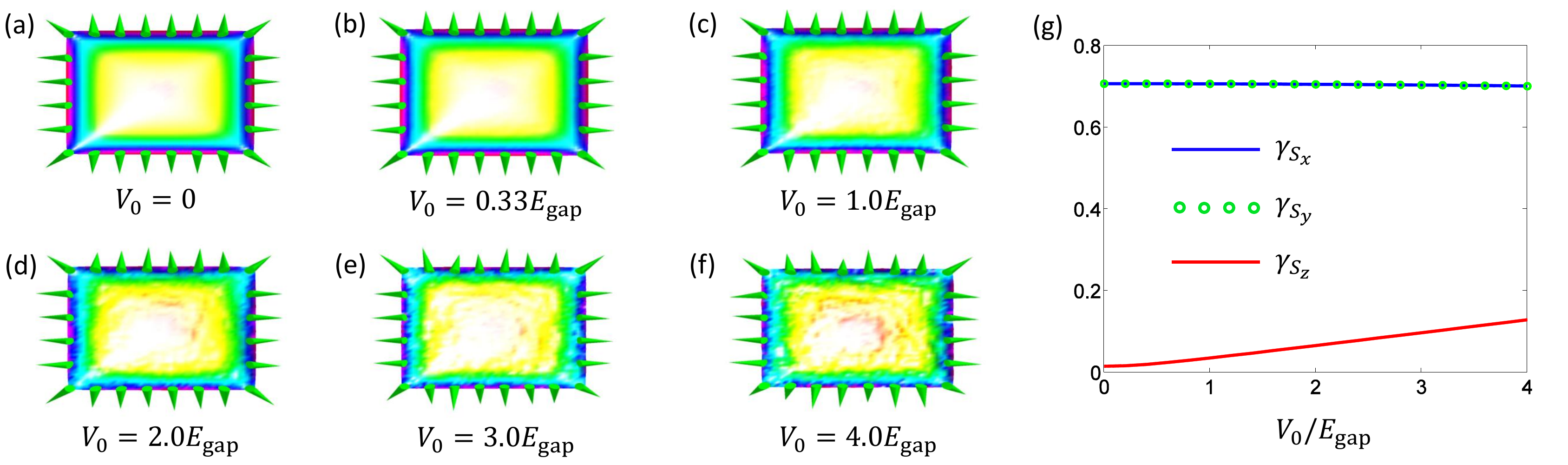}\caption{Effect of magnetic disorder scattering on the edge spin texture. The parameters for the numerical calculation are the same as those in Fig.~\ref{nonmagdisorder} $A_1=A_2=-B=1.0$ and $m_z=0.3$. (a-f) The edge spin configuration with different magnitudes of the total magnetic disorder potential $V_0$ [$=(V_x^2+V_y^2+V_z^2)^{1/2}$]; (g) Root Mean Square for $S_{x,y,z}$ versus disorder strength $V_0$. Relative to the nonmagnetic disorder regime, the magnetic disorder has a stronger effect in randomizing the wave-function distribution of edge states, but clearly has a weaker scattering effect on the spin texture.}
\label{magdisorder}
\end{figure}
The numerical results for nonmagnetic and magnetic disorder effects are shown in Fig.~\ref{nonmagdisorder} and Fig.~\ref{magdisorder}, respectively. The effects of disorder scattering on the spin textures are shown in (a) to (f) in the two figures, from which one can see that the topological spin texture is nearly unaffected even when the disorder strength equals the bulk gap $E_{\rm gap}$. In Fig.~\ref{nonmagdisorder} (g) and Fig.~\ref{magdisorder} (g) we show the Root Mean Square of $x, y$, and $z$ components of the edge spin
\begin{eqnarray}
\gamma_{S_\alpha}=\bigr[\langle S_{\alpha}^2\rangle\bigr]^{1/2}, \ \alpha=x,y,z,
\end{eqnarray}
with these spin components satisfying
\begin{eqnarray}
\langle S_{x}^2\rangle+\langle S_{y}^2\rangle+\langle S_{z}^2\rangle=1.
\end{eqnarray}
The magnitude of $\gamma_{S_z}$ quantitatively reflects the deviation of the spatial spin configuration from the in-plane topological spin texture. It can be seen that $\gamma_{S_z}$ exhibits a very weak dependence on the disorder scattering, especially in weak (magnetic and nonmagnetic) disorder regime with $V_0<0.2E_{\rm gap}$ [Fig.~\ref{nonmagdisorder} (g) and Fig.~\ref{magdisorder} (g)]. Even in the strongest disorder regime with $V_0=4.0E_{\rm gap}$, the relative Root Mean Square $p=\gamma_{S_z}/(\gamma_{S_x}+\gamma_{S_y}+\gamma_{S_z})$ is less than $12\%$ for nonmagnetic disorder and less than $9\%$ for magnetic disorder. Actually, like that the orbital chirality of the edge modes prohibits the back scattering, the spin chirality ensures that no scattering occurs between two edge modes with opposite local spin polarizations. The effect of disorder scattering is a high-order perturbation. That is, the symmetry breaking disorders induce localized states around edge which may couple to the spin degree of edge states and then lead to high-order corrections to the topological spin texture. As a result, only in the strong disorder scattering regime which has a large density of localized states around sample edges, can the spin texture be clearly affected. On the other hand, under the same disorder strength, the edge spin texture is clearly more insensitive to magnetic disorder scattering, consistent with the results in Eqs.~\eqref{eqn:SIdisorderrelation1} and~\eqref{eqn:SIdisorderrelation2}. The weak dependence of the edge spin texture on disorder scattering implies that the proposed topological spin devices in the main text are also insensitive to the local disorder perturbations.

\subsection{Effect of discrete lattice structure anisotropy}

The topological spin texture relies on the continuous approximation of the Bloch Hamiltonian, which is valid in the low-energy regime. For the realistic materials~\cite{FangZhong1,Qikun}, the bulk gap of the quantum anomalous Hall insulator is much less than the bandwidth of the system. In this case, all physics, including the topology and the edge states, can be captured by the low-energy Bloch Hamiltonian. Therefore the continuous approximation is typically well justified.

\begin{figure}[h]
\includegraphics[width=0.8\columnwidth]{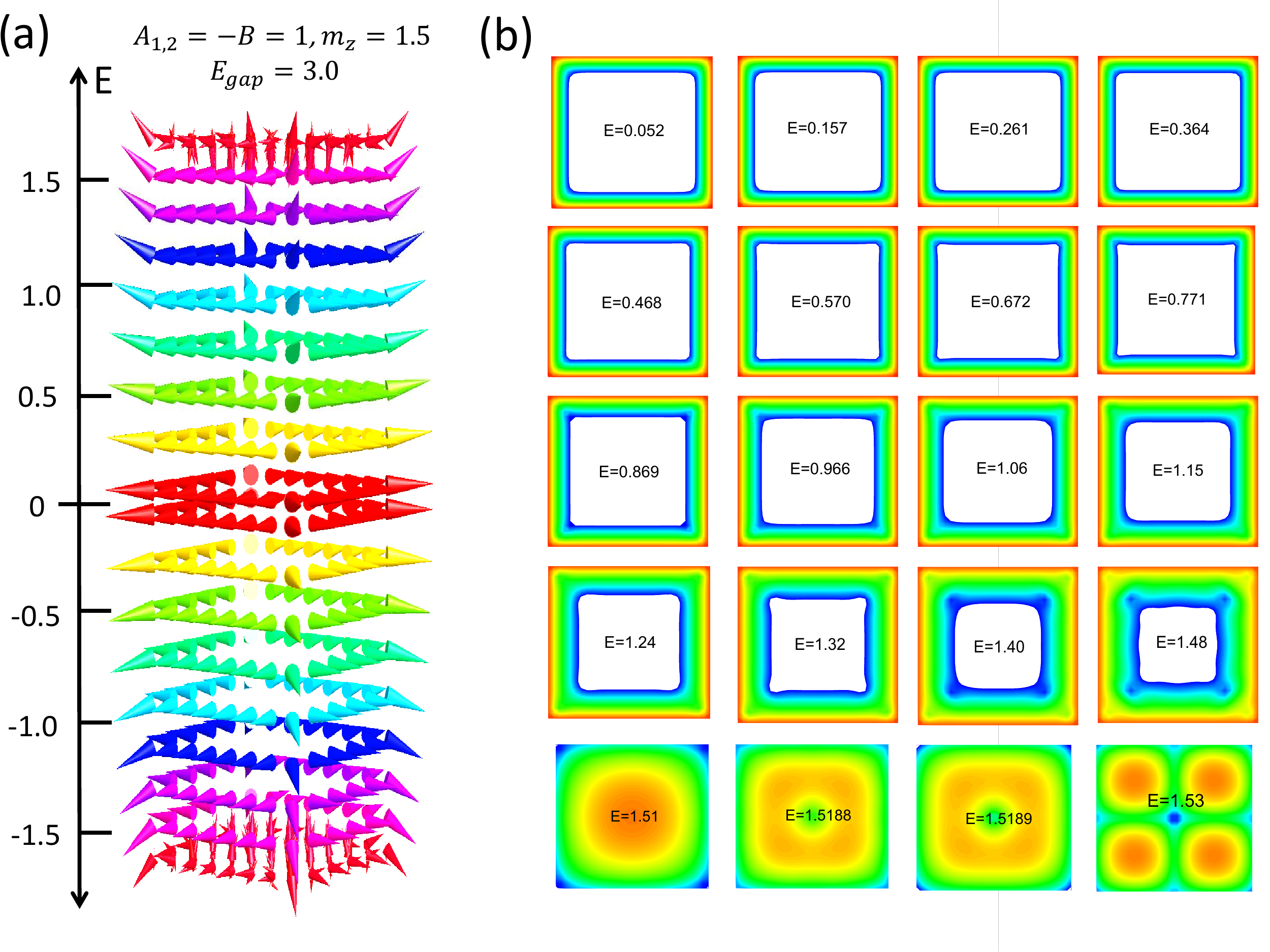}\caption{Edge spin texture byond low-energy limit in quantum anomalous Hall insulator with square boundary. The parameters are rescaled to be dimensionless that $A_1=A_2=-B=1.0$ and $m_z=1.5$, which leads to the bulk gap $E_{\rm gap}=3.0$. (a) The spin texture for edge modes of different energies, with only part of the states shown here. It can be seen that the edge states with energies $|E|<0.5$ exhibit topological spin texture, with spin polarization well within $x-y$ plane. Beyond such energy scale the edge spins exhibit a more and more pronounced tile to the perpendicular direction at square corners; (b) Wave function profiles of the edge states (for energy $E<1.5$) and bulk states (for $E>1.5$).}
\label{highenergy}
\end{figure}
The topological spin texture can in principle be scattered by the discrete lattice anisotropy in the high-energy regime. Indeed, if we consider the parameter regime that $|m_z|$ is close to or larger than $|A_{1,2}|$ and $|B|$, the bulk gap of the system is in the order of the band width. Then the high-energy edge states exist and to study them one has to go beyond the low-energy Bloch Hamiltonian. From the tight-binding Hamiltonian one can see that in the high-energy regime, due to discrete lattice anisotropy generically the Bloch Hamiltonian ${\cal H}(\bold k)$ only anticommutes with $S={\cal M}_y\sigma_x$ and $S={\cal M}_x\sigma_y$. This ensures that for a square sample with boundaries along the directions of Bravais
lattice vectors, the edge spin aligns along $x$ ($y$) axis in the edges normal to $y$ ($x$) direction and far away from sample corners. On the other hand, around the corners of the square sample the spin polarization of the high-energy edge modes may have a finite tilt to the perpendicular direction due to the broken down of the $S$ symmetry.

The numerical results are shown in Fig.~\ref{highenergy}, where we take the parameters which are rescaled to be dimensionless that $A_1=A_2=-B=1.0$ and $m_z=1.5$. In this regime the bulk band gap of the system $E_{\rm gap}=3.0$, which is close to the bandwidth. It can be seen from Fig.~\ref{highenergy} (a) that the edge states with energies $|E|<0.5$ exhibit topological spin texture, with spin polarization well within $x-y$ plane and having negligible spin tilting to $z$ direction at the corners. However, when energy increases, the edge spins at corners exhibit a more and more pronounced tile to the perpendicular direction. On the other hand, once going away from the corners by only few sites, we can find that the edge spins exactly point to $x$ or $y$ direction for all edge modes, reflecting that the chiral-like $S$ symmetry is recovered in these directions even in the high-energy regime.

It is noteworthy that for the material Bi$_2$Te$_3$, the discrete lattice anisotropy in the high energy regime is governed by $C_3$ symmetry and mirror symmetry, which can lead to $k$-cubic term $(k_+^3+k_-^3)\sigma_z$, with $k_\pm=k_x\pm ik_y$, in the Hamiltonian~\cite{Fu2009}. It can be verified directly that due to such $k$-cubic term the $S$ symmetry can only be defined in the three directions $\theta=\pi/6+m\pi/3$ ($m=0,1,2$) in the $x$-$y$ plane, reflecting the $C_3$ symmetry in such system. Then if a QAH insulator realized using magnetic Bi$_2$Te$_3$ material is cut along these directions, the edge states will be in-plane polarized, while in the corner the spin can be tilted to perpendicular direction, similar as the above results obtained in the square lattice model.

\section{Orbital angular momentum fractionalization}

The edge channel of the quantum anomalous Hall insulator can be described by 1D chiral Luttinger liquid. Furthermore, the topological spin texture leads to quantized Berry phases, which define nontrivial topological states in the boundary. Taking into account the Berry phase effect, the chiral edge states can be governed by the following effective Hamiltonian
\begin{eqnarray}\label{eqn:SILuttinger}
H_{\rm edge}=i v_{\rm edge}\int d\tilde x\psi_s^*(\tilde x)\bigr[\partial_{\tilde x}-i{\cal A}_s(\tilde x)\bigr]\psi_s(\tilde x).
\end{eqnarray}
Here $\psi_s$ denotes the orbital wave-function of the edge states. The integral of ${\cal A}_s$ along the 1D boundary gives $\oint d\tilde x{\cal A}_s(\tilde x)={\cal N}_{1d}\pi$. The $\pi$-Berry phase is equivalent to a half magnetic flux-quanta threading through the 2D sample and encircled by the edge. According to the study by Wilczek in 1982~\cite{Wilczek}, a half quantum flux can lead to $1/2$-fractionalization of the orbital angular momentum. As a result, for the 2D sample with circular geometry, the orbital angular momentum of the edge modes should be fractionalized as
$l_z=m+{\cal N}_{1d}/2$, with $m$ being integers. The energy spectrum of the edge states is corresponding to fractionalization of the orbital angular momentum and is given by
\begin{eqnarray}
{\cal E}_{l_z}=\bigr(m+\frac{1}{2}{\cal N}_{1d}\bigr)\frac{v_{\rm edge}}{R_{l_z}}, \ m=0,\pm1,\pm2,...
\end{eqnarray}
where $R_{l_z}$ is the effective radius of edge state wave function. Due to the $1/2$-fractionalization no zero-energy edge state exists. Then the number of edge states is $N=even$. This result can also be derived from the Eq.~\eqref{eqn:SIeigenstate3}. If separating the edge state wave-function by spin and orbital parts $|\phi^{\rm edge}_{m}(r,\varphi)\rangle=\phi_m(r)|\chi_s(\varphi)\rangle$, from Eq.~\eqref{eqn:SIeigenstate3} one can obtain the energy by
\begin{eqnarray}
{\cal E}_m&=&2A_1\langle\phi_m(r)|\frac{1}{r}|\phi_m(r)\rangle\langle\chi_s(\varphi)|\sigma_r(m-i\partial_\varphi)|\chi_s(\varphi)\rangle\nonumber\\
&=&\bigr(m+\frac{1}{2}\bigr)\frac{2A_{1}}{R_m}, \ m=0,\pm1,\pm2,...
\end{eqnarray}
where $R_m$ is the expectation value of $1/r$. Now let us thread an additional magnetic flux $\Phi$ through the sample, which is described by a gauge $\bold A=A_{\varphi}\hat e_\varphi$, with $A_\varphi=(\Phi/\Phi_0)\frac{1}{r}$, where $\Phi_0$ represents the flux quanta. In the presence of the external magnetic flux, the equation for $|\phi^{\rm edge}_{m}(r,\varphi)\rangle$ reads
\begin{eqnarray}
\biggr\{\bigr[2B(\frac{1}{r}\partial_rr\partial_r-\frac{(m-\Phi/\Phi_0)^2-i2(m-\Phi/\Phi_0)\partial_{\varphi}-\partial_{\varphi}^2}{r^2})+m_z\bigr]\sigma_z -i2A_{2}\sigma_{\varphi}\partial_{r}-\frac{2A_{1}}{r}\sigma_{r}(m&-&\frac{\Phi}{\Phi_0}-i\partial_\varphi)\biggr\}|\phi^{\rm edge}_{m}(r,\varphi)\rangle\nonumber\\
&&={\cal E}_m|\phi^{\rm edge}_{m}(r,\varphi)\rangle.
\end{eqnarray}
From the above equation one can find that the energy spectrum of edge states is shifted to be $\frac{2A_{1}}{R_m}(m+1/2-\Phi/\Phi_0)$.
Therefore, when an additional magnetic $1/2$-flux-quanta $\Phi=\Phi_0/2$ is threaded through the sample, the edge state $|\phi^{\rm edge}_{0}(r,\varphi)\rangle$ is exactly pushed to zero energy, and the number of edge states becomes $N=odd$. The change in the number of edge modes between {\it even} and {\it odd} by threading a half flux-quanta provides an observable for the $1/2$-fractionalization of orbital angular momentum.



\noindent


\begin{thebibliography}{99}

\bibitem{QHE} K. V. Klitzing, G. Dorda, and M. Pepper, Phys. Rev. Lett. {\bf 45}, 494 (1980).

\bibitem{FQHE} D. C. Tsui, H. L. Stormer, and A. C. Gossard, Phys. Rev. Lett.
{\bf 48}, 1559 (1982).

\bibitem{TKNN} D. J. Thouless, M. Kohmoto, M. P. Nightingale, and M. den Nijs, Phys. Rev. Lett. {\bf 49}, 405 (1982).

\bibitem{Haldane} F. D. M. Haldane, Phys. Rev. Lett. {\bf 61}, 2015 (1988).

\bibitem{TI1} M. Z. Hasan and C. L. Kane, Rev. Mod. Phys. {\bf 82}, 3045 (2010).

\bibitem{TI2} X. -L. Qi and S. -C. Zhang, Rev. Mod. Phys. {\bf 83}, 1057 (2011).

\bibitem{QAHE1}  X. -L. Qi, Y. -S. Wu, and S. -C. Zhang, Phys. Rev. B {\bf 74}, 085308 (2006).

\bibitem{QAHE2} C. -X. Liu,  X. -L. Qi, X. Dai, Z. Fang, and S. -C. Zhang, Phys. Rev. Lett. {\bf 101}, 146802 (2008).

\bibitem{Congjun2008} C. Wu, Phys. Rev. Lett. {\bf 101}, 186807 (2008).

\bibitem{FangZhong1} R. Yu, {\it et al.,} Science, {\bf 329}, 61 (2010).

\bibitem{QAHE3} X. -J. Liu, X. Liu, C. Wu, and J. Sinova, Phys. Rev. A {\bf 81}, 033622 (2010).

\bibitem{QAHE4} K. Nomura and N. Nagaosa, Phys. Rev. Lett. {\bf 106}, 166802 (2011).

\bibitem{Chaoxing} X. Liu, H. -C. Hsu, and C. -X. Liu, Phys. Rev. Lett. {\bf 111}, 086802 (2013).

\bibitem{Qikun} C. -Z. Chang {\it et al.,} Science, {\bf 340}, 167 (2013).

\bibitem{3DTI1} H. Zhang, C. -X. Liu, X.-L. Qi, X. Dai, Z. Fang, and S.-C. Zhang, Nature Phys. {\bf 5}, 438 (2009).

\bibitem{3DTI2} Y. Xia, et al., Nature Phys. {\bf 5}, 398 (2009).

\bibitem{Qi2008} X. -L. Qi,T. L. Hughes, and S. -C. Zhang, Phys. Rev. B
{\bf 78}, 195424 (2008).

\bibitem{Liu2013} X. -J. Liu, K. T. Law, and T. K. Ng, Phys. Rev. Lett. {\bf 112}, 086401 (2014).

\bibitem{Liu2009} X. -J. Liu,  M. F. Borunda, X. Liu, and J. Sinova, Phys. Rev. Lett. {\bf 102}, 046402 (2009).

\bibitem{Lin} Y. -J. Lin, K. Jim\'{e}nez-Garc\'{i}a, and  I. B. Spielman, Nature {\bf 471}, 83 (2011).

\bibitem{Wang} P. Wang, Z.-Q. Yu, Z. Fu, J. Miao, L. Huang, S. Chai, H. Zhai, and J. Zhang, Phys. Rev. Lett. {\bf 109}, 095301 (2012).

\bibitem{MIT} L. W. Cheuk, A. T. Sommer, Z. Hadzibabic, T. Yefsah, W. S. Bakr, and M. W. Zwierlein, Phys. Rev. Lett. {\bf 109}, 095302 (2012).

\bibitem{Liu2013AIII} X. -J. Liu, Z. -X. Liu, and M. Cheng, Phys. Rev. Lett. {\bf 110}, 076401 (2013).

\bibitem{SI} See Supplementary Material foe more details.

\bibitem{Schnyder} A. P. Schnyder, S. Ryu,  A. Furusaki, and A. W. W. Ludwig, Phys. Rev. B {\bf 78}, 195125 (2008).

\bibitem{Halperin} B. I. Halperin, Phys. Rev. B {\bf 25}, 2185 (1982).

\bibitem{Wen1990} X. -G. Wen, Phys. Rev. B {\bf 41}, 12838 (1990).

\bibitem{Wilczek} F. Wilczek, Phys. Rev. Lett. {\bf 48}, 1144 (1982).

\bibitem{Sankar2004} I. \v{Z}uti\'{c}, J. Fabian, S. Das Sarma, Rev. Mod. Phys. {\bf 76}, 323 (2004).

\bibitem{Julliere} M. Julliere, Phys. Lett. {\bf 54}, 225 (1975).

\bibitem{lee1981} D. S. Fisher and P. A. Lee, Phys. Rev. B {\bf 23}, 6851 (1981).

\end{thebibliography}
\end{document}